# Carbon isotope effect in superconducting MgCNi$_3$

T. Klimczuk[1,2] and R.J. Cava[1]

[1]Department of Chemistry, Princeton University, Princeton NJ 08544,

[2]Faculty of Applied Physics and Mathematics, Gdańsk University of Technology,

Narutowicza 11/12, 80-952 Gdańsk, Poland,

The effect of Carbon isotope substitution on Tc in the intermetallic perovskite superconductor MgCNi$_3$ is reported. Four independent groups of samples were synthesized and characterized. The average Tc for the $^{12}$C samples was found to be 7.12±0.02 K and the average Tc for the $^{13}$C samples was found to be 6.82±0.02 K. The resulting carbon isotope effect coefficient is $\alpha_C$ = 0.54±0.03. This indicates that carbon-based phonons play a critical role in the presence of superconductivity in this compound.

PACS numbers: 74.70Ad, 74.62Bf

MgCNi$_3$ has a non-oxygen based perovskite structure and is superconducting near 7K [1]. MgCNi$_3$ and the layered borocarbides (RENi$_2$B$_2$C) [2] form a family of Ni-rich intermetallic superconductors in which the Ni *d* states are found at the Fermi level (E$_F$), and therefore the appearance of superconductivity rather then magnetism is at first unexpected. Considering their structures and the potential magnetism, MgCNi$_3$ and the borocarbides appear to be a bridge between conventional, intermetallic superconductors and high-$T_C$ cuprates, and thus questions about the nature of their superconductivity are of interest.

For the case of MgCNi$_3$, conventional phonon-mediated, moderate [1, 3] and strong [4, 5] coupling superconductivity have been proposed from analysis of the specific heat. Strong coupling is also supported by thermopower measurements [6] and the presence of a large energy gap [4]. Further, s-wave pairing has been proposed from tunneling spectroscopy [7], point contact tunneling spectroscopy [8], $^{13}$C NMR studies [9], and specific heat measurements [3]. However the closeness of ferromagnetism [10-15], a *p*-wave superconducting order parameter and triplet superconductivity have been proposed as possibilities [15]. Finally, evidence for unconventional superconductivity has been provided by penetration depth [16], tunnelling spectra [4], and critical current measurements [17].

The Carbon atom occupies a special position in the crystal structure of MgCNi$_3$, in the center of the Ni octahedron. The Ni provides the peak in the density of states at E$_F$. From previous work, it is known that variation of the carbon stoichiometry in MgC$_{1-x}$Ni$_3$ decreases T$_C$, and that for x about 0.2 superconductivity is no longer observed [5, 18]. The important role of Carbon for the presence of superconductivity, its position in the crystal structure, and the fact that it is the lowest mass element in the compound, make Carbon a good candidate for a possible superconducting isotope effect. In this letter, we report the first study of the Carbon isotope effect in MgCNi$_3$. Surprisingly, the isotope effect is found to be substantial, suggesting that C-based phonons play an essential role in the superconducting mechanism.

Four independent groups of 0.2g samples of Mg$^{12}$CNi$_3$ and Mg$^{13}$CNi$_3$ were synthesized. In each group, one $^{12}$C and one $^{13}$C sample were made. The starting materials were bright Mg flakes (99% Aldrich Chemical), fine Ni powder (99.9% Johnson Matthey and Alpha Aesar), $^{12}$C glassy carbon spherical powder (Alfa Aesar), and amorphous $^{13}$C (99% Cambridge Isotope Laboratories, Inc.). Previous studies on MgCNi$_3$ have indicated the need to employ excess magnesium and carbon in the synthesis in order to obtain optimal carbon content [1, 5, 18]. The excess Mg is mainly vaporized during the course of the

| Series # | $T_C$ ($^{12}C$) | $T_C$ ($^{13}C$) | $\Delta T_C$ | $\alpha_C$ |
|---|---|---|---|---|
| 1 | 7.11 K | 6.83 K | 0.28 K | 0.50 |
| 2 | 7.14 K | 6.84 K | 0.30 K | 0.54 |
| 3 | 7.11 K | 6.80 K | 0.31 K | 0.56 |
| 4 | 7.13 K | 6.82 K | 0.31 K | 0.55 |
| average | 7.12 ± 0.02 K | 6.82 ± 0.02 K | 0.300 ± 0.014 K | 0.54 ± 0.03 |

reaction, and any excess carbon is present as elemental graphite [1] in the final product. After thorough mixing, the starting materials were pressed into pellets, wrapped in Zirconium foil, placed on an $Al_2O_3$ boat, and fired in a quartz tube furnace under a 95% Ar / 5% $H_2$ atmosphere. The initial furnace treatment began with a half hour at 600°C, followed by 1 hr at 900°C. After cooling, the samples were reground, pressed into pellets, and placed back in the furnace under identical conditions at 900°C. The latter step was repeated three additional times. Following the final heat treatment, the samples were analyzed with powder X-ray diffraction using $CuK_\alpha$ radiation. The superconducting transition temperature was determined by zero field cooling AC magnetization ($H_{DC}$=5 Oe, $H_{AC}$=3 Oe, f=10 kHz) from 1.9K to 8K (PPMS – Quantum Design).

The crystallographic cell parameter was determined from least-squares fits to 9 X-ray reflections between 20 and 90 degrees 2θ for representative $^{12}C$ and $^{13}C$ samples. There is no observable isotope mass dependence of the unit cell parameter: a=3.8109(6) and a=3.8116(15) for $Mg^{12}CNi_3$ and $Mg^{13}CNi_3$ respectively. These lattice parameters are in agreement with previously reported values for stoichiometric $MgCNi_3$ [1, 18, 20].

Zero field cooled normalized magnetization data are presented in Figure 1 for samples containing both $^{12}C$ and $^{13}C$ isotopes: data from the $Mg^{12}CNi_3$ samples are plotted in the top panel (closed squares) and data from the $Mg^{13}CNi_3$ samples are plotted in the bottom panel (closed circles). As can be seen from a cursory inspection of Figure 1, there is a clear shift of the superconducting critical temperature ($T_C$) to lower temperatures in the $^{13}C$ samples. The magnetization data in the plot have been normalized to the 4.5K values to facilitate easy comparison near Tc. For a detailed comparison, $T_C$ was taken as the temperature where the extrapolation of the steepest slope of the real part of the diamagnetic magnetization versus temperature curves intersects the extrapolation of the normal state magnetization to lower temperatures, removing ambiguities inherent in the use of an "onset" temperature for defining Tc. This extrapolation for one representative sample is presented in the bottom panel of Figure 1.

The Tc data are summarized in Table 1. The average shift, $\Delta T_C = T_C(^{12}C) - T_C(^{13}C)$, was calculated to be 300mK, with the standard deviation value 15mK. The Carbon isotope effect ($\alpha_C$) for each of the four groups was then calculated from the following relation:

$$\alpha_C = -\frac{\Delta \ln(T_C)}{\Delta \ln(M_C)} = -\frac{\ln\frac{T_C(^{12}C)}{T_C(^{13}C)}}{\ln\frac{12}{13}}$$, where $T_C$ and $M_C$ are

a critical temperature and a molecular isotope mass ratio, respectively.

The average value of the carbon isotope effect is therefore $\alpha_C = 0.54 \pm 0.03$. Table 1 presents the results for all sample groups.

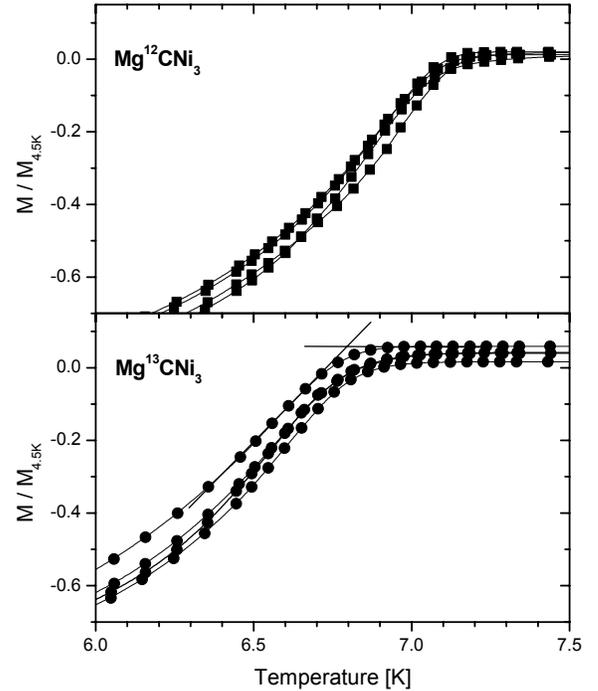

**Fig. 1** AC normalized magnetization (M / $M_{4.5K}$) data for all samples in the temperature range 6 – 7.5K. Upper panel presents $Mg^{12}CNi_3$ (closed squares) and bottom panel presents $Mg^{13}CNi_3$ data (closed circles).

For a phonon – mediated superconductor, McMillan has shown that $T_C$ is proportional to $M^{-\alpha}$ ($T_C \propto \frac{1}{M^\alpha}$), where α is given by [19]:

$$\alpha = \frac{1}{2}\left(1 - \left(\mu^* \ln\frac{\Theta_D}{1.45 T_C}\right)^2 \frac{1+0.62\lambda}{1+\lambda}\right).$$

The Debye temperature ($\Theta_D$) values reported for $MgCNi_3$ range between 256 and 351 K [3-5, 20]. The electron-phonon coupling constant (λ) has been reported to be in the range from λ=0.77 [1], 0.83 [3] (from specific heat measurements) to λ(0)=1.4 [6] (from thermopower measurements). We take the values $\Theta_D$ = 256K and λ = 0.77 for our calculations, and note that α is not very sensitive to these parameters. $\alpha_C$ is sensitive to the Coulomb coupling constant between electrons ($\mu^*$) however, which has been reported to be in the range 0.10 – 0.15 [21]. This yields an expected $\alpha_C$ in the range of 0.4 – 0.45 from the general characterization of $MgCNi_3$. Our observed value of 0.54±0.03 for the isotope effect of carbon is higher than but similar to that value.

To summarize we have measured the Carbon isotope effect in $MgCNi_3$. The obtained value, $\alpha_C = 0.54 \pm 0.03$, is very close to what is expected for a phonon mediated superconductor. This result indicates that carbon-based phonons play a critical role in the presence of superconductivity in this compound, which shows some unconventional characteristics in its superconducting state.

The Carbon isotope effect $\alpha_C = 0.54$ is higher but similar to that reported for $Rb_3C_{60}$ $\alpha_C$=0.37 [22], 0.21 [23], and substantially higher than that for the related borocarbide $YNi_2B_2C$, where $\alpha_C$ is reported to be very close to zero ($\alpha_C$=0.07) [24]. For the borocarbide, however, the Boron isotope effect is measured to be $\alpha_B$=0.25 [24] and 0.21 [25] indicating a substantial contribution of boron derived phonons to the superconducting mechanism. The boron isotope effect value in the borocarbide is very close to that in $MgB_2$, reported to be in the range $\alpha_B$=0.28 [26] to $\alpha_B$=0.30 [27]. The implication of our measurement is that the vibration of the Carbon within the $Ni_6$ octahedron strongly modulates the charge in the Ni-Ni near neighbour bonds that dominate the density of states at the Fermi energy ($E_F$). If so, then this would be similar to the case in the borocarbide superconductors, where B-C-B phonons have been proposed to strongly influence the charge distribution at $E_F$ in states dominated by Ni orbital character [28].

We thank M. Mierzejewski and M. Maska (University of Silesia, Poland) for fruitful discussion. This work was supported by the US Department of Energy, grant DE-FG02-98-ER45706.